# H-atom addition and abstraction reactions in mixed CO, $H_2CO$ and $CH_3OH$ ices – an extended view on complex organic molecule formation

K.-J. Chuang,[1,2]⋆ G. Fedoseev,[1] S. Ioppolo,[3] E. F. van Dishoeck[2] and H. Linnartz[1]

[1]*Sackler Laboratory for Astrophysics, Leiden Observatory, Leiden University, PO Box 9513, NL-2300 RA Leiden, the Netherlands*
⋆E-mail: chuang@strw.leidenuniv.nl
[2]*Leiden Observatory, Leiden University, PO Box 9513, NL-2300 RA Leiden, the Netherlands*
[3]*Department of Physical Sciences, The Open University, Walton Hall, Milton Keynes MK7 6AA, UK*

**ABSTRACT**

Complex organic molecules (COMs) have been observed not only in the hot cores surrounding low- and high-mass protostars, but also in cold dark clouds. Therefore, it is interesting to understand how such species can be formed without the presence of embedded energy sources. We present new laboratory experiments on the low-temperature solid state formation of three complex molecules – methyl formate ($HC(O)OCH_3$), glycolaldehyde ($HC(O)CH_2OH$) and ethylene glycol ($H_2C(OH)CH_2OH$) – through recombination of free radicals formed via H-atom addition and abstraction reactions at different stages in the $CO \rightarrow H_2CO \rightarrow CH_3OH$ hydrogenation network at 15 K. The experiments extend previous CO hydrogenation studies and aim at resembling the physical–chemical conditions typical of the CO freeze-out stage in dark molecular clouds, when $H_2CO$ and $CH_3OH$ form by recombination of accreting CO molecules and H-atoms on ice grains. We confirm that $H_2CO$, once formed through CO hydrogenation, not only yields $CH_3OH$ through ongoing H-atom addition reactions, but is also subject to H-atom-induced abstraction reactions, yielding CO again. In a similar way, $H_2CO$ is also formed in abstraction reactions involving $CH_3OH$. The dominant methanol H-atom abstraction product is expected to be $CH_2OH$, while H-atom additions to $H_2CO$ should at least partially proceed through $CH_3O$ intermediate radicals. The occurrence of H-atom abstraction reactions in ice mantles leads to more reactive intermediates (HCO, $CH_3O$ and $CH_2OH$) than previously thought, when assuming sequential H-atom addition reactions only. This enhances the probability to form COMs through radical-radical recombination without the need of UV photolysis or cosmic rays as external triggers.

## 1 INTRODUCTION

CO is the second most abundant molecule in the interstellar medium (ISM) after $H_2$ (Ohishi, Irvine & Kaifu 1992). It is formed in the gas phase and despite its high volatility, carbon monoxide accretes on the surfaces of grains in the dense and cold parts of molecular clouds. After water, CO is the second most abundant component of interstellar ices (Pontoppidan 2006). Observational data show that rather than mixing with $H_2O$, the bulk of the CO accretes on top of a previously formed $H_2O$-rich polar ice, forming an apolar CO- rich ice layer (Tielens et al. 1991, Öberg et al. 2011a; Boogert & Ehrenfreund 2004; Gibb et al. 2004; Mathews et al. 2013, Boogert, Gerakines & Whittet 2015). The resulting CO-coating, in turn, is thought to react with impacting H-atoms, producing $H_2CO$ via the HCO intermediate radical and subsequently $CH_3OH$ through $CH_3O$ or, possibly, $CH_2OH$ radical intermediates. This surface formation route is generally considered to be the chemical pathway explaining the observed abundance of methanol in dense clouds, both in the solid state and in the gas phase. The process has been subject of numerous experimental (Hiraoka et al. 1994; Zhitnikov & Dmitriev 2002; Watanabe & Kouchi 2002; Fuchs et al. 2009), theoretical and modelling (Tielens & Hagen 1982; Shalabiea & Greenberg 1994; Cuppen et al. 2009, Vasyunin & Herbst 2013) studies. For a recent review, see also Linnartz, Ioppolo & Fedoseev (2015). Moreover, combined laboratory and observational data show that CO and $CH_3OH$ are intimately mixed in interstellar ices. This is fully consistent with a common chemical history (Cuppen et al. 2011).

Solid methanol, in turn, has been proposed as a starting point for the formation of complex organic molecules (COMs). Experiments involving energetic processing, e.g., UV photolysis (Öberg et al. 2009; Henderson & Gudipati 2015), soft X-ray irradiation (Chen et al. 2013), high-energy electron (Bennett et al. 2007; Maity, Kaiser & Jones 2015) and ion (Moore, Ferrante & Nuth 1996; de Barros et al. 2011) bombardment, and low-energy electron radiolysis (Boamah et al. 2014) of solid methanol ice, show that COMs form upon recombination of dissociation products. This is in line with a number of astronomical observations (for recent reviews see Herbst & van Dishoeck 2009; Caselli & Ceccarelli 2012). However, these experiments do not explain the recent detection of COMs in dark clouds where icy grains are not exposed to strong UV fields or have been heated. In particular, methyl formate, acetaldehyde, dimethyl ether and ketene have been detected in the cold pre-stellar core L1689B (Bacmann et al. 2012). The same species have also been seen in the cold pre-stellar core B1-b (Cernicharo et al. 2012), in the cold outer envelopes of low-mass protostars (Öberg et al. 2010), and in outflow spots in dark clouds where the ice mantles are liberated by shocks (Arce et al. 2008; Öberg et al. 2011b). Acetaldehyde and ketene were also detected in the cold pre-stellar core L1544 (Vastel et al. 2014). These observations clearly hint for a scenario in which COMs also form at temperatures below 15 K. Recent work by Fedoseev et al. (2015) demonstrated *non-energetic* routes to form COMs by surface hydrogenation of CO molecules.

An efficient pathway, creating a C–C backbone without the involvement of energetic processing, has the potential to form COMs earlier than expected during the chemical evolution of interstellar clouds and moreover will increase the chemical diversity. The solid state formation of various two- or even three-carbon bearing species was already suggested in a number of astrochemical models. Charnley, Rodgers & Ehrenfreund (2001) and Charnley & Rodgers (2005) studied C-atom addition reactions to HCO and HCCO radicals, whereas Garrod, Weaver & Herbst (2008) and Woods et al. (2012) worked on a backbone extension through recombination of various carbon-bearing intermediates, such as $CH_3$, HOCO and the CO hydrogenation intermediates; HCO, $CH_3O$ and $CH_2OH$. Some of these reactions were already proposed quite some time ago, i.e., by Agarwal et al. (1985) and Schutte (1988), to explain results obtained after energetic processing of ices. Up to now, however, experimental studies verifying the formation of COMs along non-energetic pathways have been largely lacking, with exception of the above-mentioned study by Fedoseev et al. (2015) who showed that glycolaldehyde (HC(O)$CH_2OH$) and ethylene glycol ($H_2C(OH)CH_2OH$) form at low temperatures in CO+H deposition experiments. In their study, Fedoseev and co-workers combined molecule specific desorption temperatures and ionization fragmentation patterns for the newly formed species to conclude that recombination of HCO radicals yields glyoxal (HC(O)CHO) that is subsequently converted to glycolaldehyde and ethylene glycol upon two or four consecutive H-atom additions, respectively. They also suggested a possible formation route of methyl formate (HC(O)OCH$_3$) through recombination of HCO and CH$_3$O radicals, but it was not possible to experimentally confirm this.

The concept of non-energetic H-atom abstraction reactions is not new. Tielens & Hagen (1982) presented modelling calculations that emphasized the importance of H-atom abstraction reactions in grain-surface hydrogenation sequences. Experimental work by Nagaoka, Watanabe & Kouchi (2005) investigated the exposure of methanol ice to D-atoms. They found that partially deuterated methanol (CH$_2$DOH, CHD$_2$OH and CD$_3$OH) is quickly formed upon D-atom exposure of solid CH$_3$OH at 10 K. Following results from ab initio calculations, the authors proposed that one of the possible H–D substitution pathways takes place via H-atom abstraction from the methyl side of methanol to form the hydroxymethyl radical (CH$_2$OH), after which D-atom addition forms CH$_2$DOH. Subsequently, Hidaka et al. (2009) proposed H–D substitution in H$_2$CO to also originate from an H-atom abstraction yielding HCO followed by D-atom addition to form HDCO. In both studies, the existence of abstraction reactions is crucial to explain the formation of deuterated molecules. Similar H-atom abstraction reactions may be triggered by H-atoms as well, i.e. instead of D-atoms, which would effectively increase the total amount of HCO, CH$_3$O and CH$_2$OH radicals formed in the ice, increasing the probability that recombination reactions result in COM formation.

The aim of this study is to verify the formation of COMs through H$_2$CO+H and CH$_3$OH+H, extending on Fedoseev et al. (2015) by focusing on the influence of H-atom induced abstraction reactions along with H-atom addition events. We experimentally investigate hydrogenation of pure ice samples (H$_2$CO or CH$_3$OH) as well as ice mixtures (H$_2$CO+CO, CH$_3$OH+CO, or H$_2$CO+CH$_3$OH). Both the existence of abstraction reactions involving H$_2$CO and CH$_3$OH and the possible formation of COMs are discussed. Special care is taken to verify that the COMs are products formed during codeposition, and not the result of thermally induced chemistry upon heating or due to contaminations.

## 2 EXPERIMENTAL PROCEDURE

### 2.1 Description of the setup

All experiments are performed under ultra-high vacuum (UHV) conditions, using the SURFRESIDE$^2$ setup that has been described in detail by Ioppolo et al. (2013). This setup consists of three distinct UHV chambers: a main chamber with base pressure of $\sim 10^{-10}$ mbar and two beam line chambers with base pressures in the range of $10^{-9}$–$10^{-10}$ mbar. These chambers are separated by shutters, allowing independent operation of the individual parts. In the main chamber, a rotatable gold-coated copper substrate is mounted on the tip of a cold head of a closed-cycle helium cryostat. Accessible temperatures range from 13 to 300 K and ice samples are deposited on the substrate with monolayer precision (where 1 ML is assumed to be $\sim 1 \times 10^{15}$ molecules cm$^{-2}$). The absolute temperature is accurate to better than 2 K, while the relative precision is better than 0.5 K.

The beam line chambers comprise different atom sources: a Hydrogen Atom Beam Source (HABS, Dr. Eberl MBE-Komponenten GmbH; see Tschersich 2000) generating H- or D-atoms by thermal cracking $H_2$ and $D_2$, and a Microwave Atom Source (MWAS, Ox- ford Scientific Ltd; see Anton et al. 2000) capable of producing H-, D-, O-, or N-atoms as well as various radicals by cracking selected parent molecules in a capacitively coupled microwave discharge (275 W at 2.45 GHz). Here only the MWAS is used, as the HABS chamber may contain CO contaminations when operated at high temperatures of the tungsten filament. Typical atom fluxes amount to roughly $6 \times 10^{14}$ atoms $min^{-1} cm^{-2}$. Along the path of both beam lines, a nose-shape quartz pipe is placed behind the shutter to efficiently quench excited atoms and non-dissociated molecules through collisions with the wall of the pipe. In addition, two separate dosing lines in the main chamber are used for deposition of molecular components of the ice, i.e. CO, $H_2CO$ and $CH_3OH$. The individual gas samples are prepared by introducing single gases into a distinct well prepumped ($<1 \times 10^{-4}$ mbar) full-metal reservoir. By means of a high-precision full-metal leak valve gas vapours are introduced into the UHV chamber with normal and 68° incidence angles to the sample. Degassed liquid $CH_3OH$ (Sigma- Aldrich 99.9 per cent) is used to obtain $CH_3OH$ vapour and solid

**Table 1.** Overview of the performed experiments.

| No. | Experiment | $T_{sample}$ (K) | Ratio | $Flux_{Molecule}$ ($min^{-1}cm^{-2}$) | | $Flux_H$ ($min^{-1}cm^{-2}$) | Time (min) | | | |
|---|---|---|---|---|---|---|---|---|---|---|
| Hydrogenation of pure $H_2CO$ and $CH_3OH$ | | | | | | | | | | |
| 1.1 | $H_2CO+H$ | 15 | 1:30 | $2 \times 10^{13}$ | | $6 \times 10^{14}$ | 360 | | | |
| 1.2 | $H_2CO$ | 15 | | $2 \times 10^{13}$ | | | 360 | | | |
| 1.3 | H | 15 | | | | $6 \times 10^{14}$ | 360 | | | |
| 1.4 | no deposition | 15 | | | | | 360 | | | |
| 1.5 | $CH_3OH+H$ | 15 | 1:30 | $2 \times 10^{13}$ | | $6 \times 10^{14}$ | 360 | | | |
| 1.6 | $CH_3OH$ | 15 | | $2 \times 10^{13}$ | | | 360 | | | |
| 1.7 | $^{13}CH_3OH+H$ | 15 | 1:20 | $3 \times 10^{13}$ | | $6 \times 10^{14}$ | 360 | | | |
| 1.8 | $^{13}CH_3OH$ | 15 | | $3 \times 10^{13}$ | | | 360 | | | |
| 1.9 | $H_2CO+H$ | 25 | 1:30 | $2 \times 10^{13}$ | | $6 \times 10^{14}$ | 360 | | | |
| 1.10 | $H_2CO+H$ | 50 | 1:30 | $2 \times 10^{13}$ | | $6 \times 10^{14}$ | 360 | | | |
| 1.11 | $H_2CO+H$ | 15 | 1:30 | $2 \times 10^{13}$ | | $6 \times 10^{14}$ | 850 | | | |
| No. | Experiment | $T_{sample}$ (K) | Ratio | $Flux_{Molecule(1)}$ ($min^{-1}cm^{-2}$) | $Flux_{Molecule(2)}$ ($min^{-1}cm^{-2}$) | $Flux_H$ ($min^{-1}cm^{-2}$) | Time (min) | Relative abundance | Relative abundance | Relative abundance |
| Hydrogenation of CO, $H_2CO$, $CH_3OH$ mixtures | | | | | | | | $HC(O)OCH_3$ | $HC(O)CH_2OH$ | $H_2C(OH)CH_2OH$ |
| 2.1 | $CO+H_2CO+H$ | 15 | 1:1:30 | $2 \times 10^{13}$ | $2 \times 10^{13}$ | $6 \times 10^{14}$ | 360 | 0.46 | 0.15 | 0.49[a] |
| 2.2 | $CO+H_2CO$ | 15 | 1:1 | $2 \times 10^{13}$ | $2 \times 10^{13}$ | | 360 | | | |
| 2.3 | $CO+CH_3OH+H$ | 15 | 1:1:30 | $2 \times 10^{13}$ | $2 \times 10^{13}$ | $6 \times 10^{14}$ | 360 | 0.00 | 0.14 | 0.00[a] |
| 2.4 | $CO+CH_3OH$ | 15 | 1:1 | $2 \times 10^{13}$ | $2 \times 10^{13}$ | | 360 | | | |
| 2.5 | $H_2CO+CH_3OH+H$ | 15 | 1:1:30 | $2 \times 10^{13}$ | $2 \times 10^{13}$ | $6 \times 10^{14}$ | 360 | 0.56 | 0.20 | 0.13[a] |
| 2.6 | $H_2CO+CH_3OH$ | 15 | 1:1 | $2 \times 10^{13}$ | $2 \times 10^{13}$ | | 360 | | | |
| 2.7 | $H_2CO+H$ | 15 | 1:15 | $4 \times 10^{13}$ | | $6 \times 10^{14}$ | 360 | 1.00 | 1.00 | 1.00[a] |
| 2.8 | $CO+H$ | 15 | 1:15 | $4 \times 10^{13}$ | | $6 \times 10^{14}$ | 360 | 0.00 | 0.33 | 0.30[a] |
| No. | Experiment | $T_{sample}$ (K) | Ratio | $Flux_{H2CO}$ ($min^{-1}cm^{-2}$) | $Flux_{CH3OH}$ ($min^{-1}cm^{-2}$) | | Time (min) | | | |
| Control experiments | | | | | | | | | | |
| 3.1 | $H_2CO+CH_3OH$ | 15 | 1:3.5 | $17 \times 10^{13}$ | $59 \times 10^{13}$ | – | 50 | | | |
| 3.2 | $H_2CO+CH_3OH$ | 15 | 1:0.3 | $16 \times 10^{13}$ | $4 \times 10^{13}$ | – | 50 | | | |
| 3.3 | $H_2CO+CH_3OH$ | 15 | 1:0.1 | $16 \times 10^{13}$ | $2 \times 10^{13}$ | – | 50 | | | |

*Note.* All of the relative abundances are based on values in Fig. 7.

[a]These values are obtained after subtraction of the control experiment accounting for the influence of thermal processing of the ice and contaminations (i.e. red columns in the upper panel of Fig. 7).

paraformaldehyde powder (Sigma-Aldrich 95 per cent) warmed to 60–80°C under vacuum to generate $H_2CO$ vapours. Residuals are typically shorter oligomers of formaldehyde and water. A CO gas cylinder (Linde 2.0, residuals: $^{13}CO$, $N_2$ and $CO_2$) is used for the preparation of carbon monoxide containing ice samples.

The ice diagnostics are performed by using either Fourier trans- form infrared absorption spectroscopy (FT-RAIRS) or temperature- programmed desorption quadrupole mass spectrometry (TPD QMS). The first method allows *in situ* studies of species embedded, formed or consumed in the ice, but has limited sensitivity and selectivity. The FT-RAIRS covers the range between 4000 and 700 cm$^{-1}$ with a spectral resolution of 1 cm$^{-1}$. A modified Lambert–Beer's law is used to derive number densities of CO and H$_2$CO on the substrate using absorbance strength as described in Ioppolo et al. (2013). The absorbance strength values of CO, CH$_3$OH and H$_2$CO are obtained from transmission absorbance strength values as described by Watanabe et al. (2004). After completion of a codeposition and RAIRS experiment, a TPD QMS experiment is performed with a typical rate of 5 K/min to monitor thermally desorbing ice species. The TPD QMS is a more sensitive technique and combines known desorption temperatures with dissociative ionization frag-mentation patterns upon electron impact in the head of the QMS. This makes it a strong diagnostic tool to recognize newly formed species, but obviously this technique comes with the thermal processing and ultimately destruction of the ice. See also Ioppolo, Öberg & Linnartz (2014) and Linnartz, Ioppolo & Fedoseev (2015) for further technical details.

**2.2 Experimental methods**

Three distinct sets of experiments are performed and systematically listed in Table 1. Each of them addresses a specific goal.

1) Verification of CO and H$_2$CO formation by H-atom abstraction from H$_2$CO and CH$_3$OH, respectively; exps 1.1–1.11.

H-atom-induced abstraction reactions from H$_2$CO and CH$_3$OH yielding CO and H$_2$CO, respectively, are verified by codepositing pure H$_2$CO or CH$_3$OH samples and H-atoms for different settings. The applied codeposition technique has the advantage that penetration depth issues into the bulk of the ice can be circumvented, as two or more species can be deposited simultaneously rather than sequentially. The latter has been a problem in previous pre-deposition experiments (Watanabe, Shiraki & Kouchi 2003; Fuchs et al. 2009) in which only the top few layers were involved in H-atom-induced reactions. In a codeposition experiment, gas mixing ratios are easily varied and the consequent use of a high enough abundance of H-atoms compared to the molecules of interest guarantees that they all become available for encounters with H-atoms. Moreover, the formed products are trapped in the growing ice lattice and this pre-vents them from further interactions with H-atoms. Furthermore, codeposition mimics the actual processes taking place on an interstellar grain in space when the outer layer starts accreting CO molecules together with impacting H-atoms (Cuppen et al. 2009).

The newly formed species are monitored *in situ* for the full time of a 360 min codeposition experiment by means of RAIRS. After completion of the codeposition, a TPD QMS experiment is per-formed. Control experiments comprise pure H$_2$CO and CH$_3$OH depositions without H-atoms (exps 1.2 and 1.6), H-atom deposition without H$_2$CO or CH$_3$OH molecules (exp. 1.3), and a blank experiment without any deposition (exp. 1.4). The formation of H$_2$CO in CH$_3$OH+H is further verified using $^{13}$C-labelled methanol.

2) Verification of COM formation via radical–radical interactions in the aforementioned system; exps 2.1–2.8.

The hydrogenation experiments described in the previous section are further extended by performing codeposition experiments of bi-nary ice mixtures, H$_2$CO+CO, CH$_3$OH+CO and H$_2$CO+CH$_3$OH with H-atoms. In the next section, it is shown that, for these experiments, COMs can be detected, specifically methyl formate (HC(O)OCH$_3$), glycolaldehyde (HC(O)CH$_2$OH) and ethylene glycol (H$_2$C(OH)CH$_2$OH). TPD QMS is used as the main diagnostic tool to detect these COMs because of (significant) spectral overlap of the stronger vibrational modes with $H_2CO$ and $CH_3OH$ infrared absorption features (Öberg et al. 2009).

3) The final set of TPD QMS experiments (exps 3.1–3.3) is specifically designed to rule out the possibility that any COMs are found due to contaminations in the depositing gas samples or as a result of thermally induced reactions during TPD. For this purpose, a series of depositions with mixed $H_2CO:CH_3OH$ ices is followed by a regular TPD QMS experiment without any H atoms impacting. In these control experiments, the used $H_2CO:CH_3OH$ ratios are chosen to cover the entire range of values as obtained in the hydrogenation experiments after 6 h of codeposition. This allows for a systematic comparison between the TPD QMS peak intensities of $HC(O)OCH_3$, $HC(O)CH_2OH$ and $H_2C(OH)CH_2OH$ found in the hydrogenation experiments and the corresponding results from the control experiments.

# 3 RESULTS

## 3.1 CO formation upon codeposition of H and $H_2CO$

In Fig. 1, RAIRS data (exps 1.1 and 1.2) for a 15 K codeposition experiment of $H_2CO$+H (upper spectrum) and only $H_2CO$ (lower spectrum), i.e. without hydrogenation, are presented. The spectral signatures of the originally deposited $H_2CO$ are found around 1727 $cm^{-1}$ as well as at 1499 and 1253 $cm^{-1}$; an $H_2CO$ band around 1178 $cm^{-1}$ is harder to discriminate. In the H-atom addition experiment, new peaks at 1031 and 1423 $cm^{-1}$ can be assigned to the C–O stretching and O–H bending vibrational modes of $CH_3OH$, respectively (Falk & Whalley 1961). A clearly visible peak at 2138 $cm^{-1}$ is due to the stretching mode of CO. The formation of $CH_3OH$ is consistent with previous studies of successive hydrogenation of CO and $H_2CO$ (Hiraoka et al. 1994, Watanabe & Kouchi 2002; Hidaka et al. 2004; Fuchs et al. 2009). The simultaneous appearance of the 2138 $cm^{-1}$ band indicates that along with the H-atom addition reactions to formaldehyde, resulting in the formation of $CH_3OH$, also a sequence of two H-atom abstraction reactions takes place, yielding CO. Spectral features of the intermediate radicals, i.e., HCO and $CH_3O/CH_2OH$ are not observed, consistent with previous studies concluding that their abundance is low. This is due to the higher reactivity of these intermediates to H-atoms via barrierless reactions compared to the stable species that have to bypass activation barriers (Watanabe & Kouchi 2002; Fuchs et al. 2009).

The main experimental finding here is that interaction of H-atoms with $H_2CO$ molecules at 15 K not only leads to the formation of methanol but also of carbon monoxide. This is in agreement with the experimental findings of Hidaka et al. (2004).

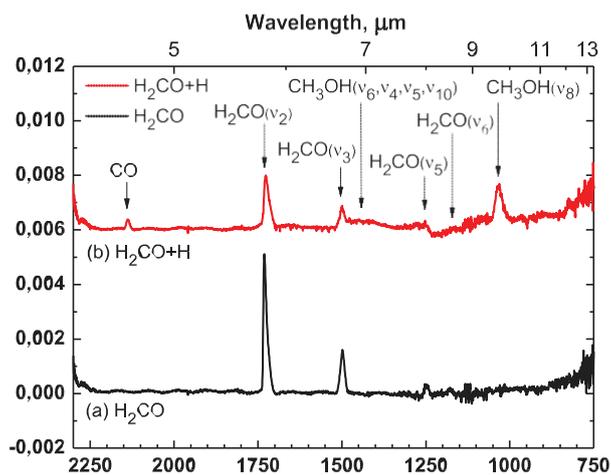

**Figure 1.** RAIR spectra obtained after 360 min (co)deposition at 15 K of (a) $H_2CO$ and (b) $H_2CO$ with H-atoms in a ratio $H_2CO:H$ = 1:30 and an H-atom flux of $6 \times 10^{14}$ atoms min$^{-1}$ cm$^{-2}$.

### 3.2 $H_2CO$ formation upon codeposition of H and $CH_3OH$

The results of a $CH_3OH+H$ 15 K codeposition experiment as well as reference data obtained for a pure $CH_3OH$ deposition (exps 1.5 and 1.6) are shown in Fig. 2. In the left-hand panel, the wavelength domain that covers the strongest (RAIR) absorption band of $H_2CO$, i.e. its C–O stretching vibration mode, is shown (see also Fig. 1); neither this band nor any other $H_2CO$ features can be found. The left shoulder of a broad band around 1652 cm$^{-1}$ could be due to the C–O stretching mode of formaldehyde, but given the absence of the other $H_2CO$ bands the full absorption feature is more likely due to polluting $H_2O$ that results from contamination in the H-atom beam. A similar band is also present in Fig. 1, but hard to see given the less accurate intensity scale used there.

More sensitive TPD QMS data are presented in the right-hand panel of Fig. 2, which confirm $H_2CO$ formation. The two strongest $m/z$ signals of $H_2CO$, i.e., $HCO^+$ (29 amu) and $H_2CO^+$ (30 amu) exhibit both a peak centred at 95 K. This desorption temperature as well as the $m/z = 30$ to $m/z = 29$ ratio of 0.6 are consistent with literature data (Fuchs et al. 2009, NIST database[1]). To further con- strain $H_2CO$ formation from $H+CH_3OH$, the experiment is repeated for $^{13}C$-labelled methanol (not shown in Fig. 2). The TPD QMS shows the same peak at 95 K, but this time the maximum has an $m/z$ value of 30 (as opposed to 29 for the regular $^{12}C$ isotope of $H_2CO$) and again a ratio of $m/z = 31$ to $m/z = 30$ of 0.6 is found. This is fully consistent with the conclusion that the species desorbing at 95 K can be assigned to formaldehyde and that it forms through hydrogen abstraction from $CH_3OH$. This means that the network derived by Hidaka et al. (2009) can be further extended to include a two-step dehydrogenation process transferring $CH_3OH$ into $H_2CO$. Also here, the intermediate $CH_2OH$ radical is not observed because of its high reactivity.

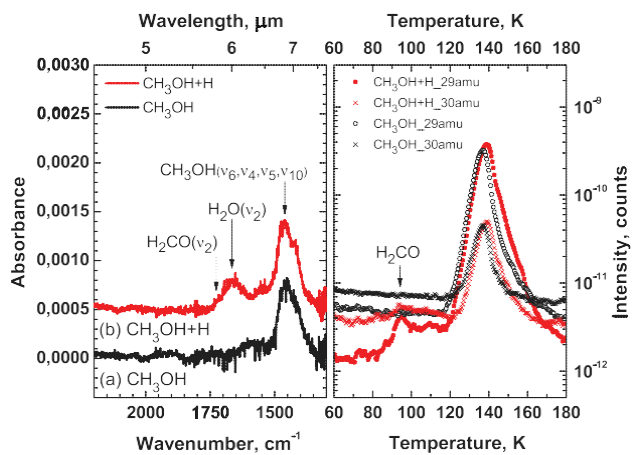

**Figure 2.** The left-hand panel shows RAIR spectra of 360 min (co)deposition at 15 K of (a) CH$_3$OH and (b) CH$_3$OH with H-atoms in a ratio CH$_3$OH:H = 1:30 and an H-atom flux of 6 × 10$^{14}$ atoms min$^{-1}$cm$^{-2}$. The shown wavelength domain corresponds to the region where the strongest absorption features of H$_2$CO should be visible. The right-hand panel presents parts of the TPD QMS spectra obtained after these two experiments for $m/z$ = 29 and 30 amu. Peaks at T∼95 and 140 K correspond to formaldehyde and methanol desorption, respectively.

### 3.3 Formation of COMs: methyl formate, glycolaldehyde and ethylene glycol

In Fig. 3, the TPD QMS spectra (exps 1.1 and 1.2) are presented for a temperature range from 15 to 225 K taken after 360 min of H$_2$CO+H codeposition and H$_2$CO deposition at 15 K, respectively. In addition to the desorption peaks from the originally deposited H$_2$CO, its hydrogenation product CH$_3$OH and its abstraction product CO (not shown), there are three more desorption peaks showing up. The TPD QMS spectra provide two ways to identify the desorbing species; via their desorption temperature and via their dissociative ionization fragmentation pattern upon electron impact in the head of the QMS. The first sublimation peak is centred at 120 K, the centre of the second one is located around 160 K, and the last one is around 200 K. Based on available data reported by Öberg et al. (2009) and recent work by Fedoseev et al. (2015), the aforementioned three desorption temperatures can be attributed to methyl formate, glycolaldehyde, and ethylene glycol, respectively.

The assignments are further supported by the fragmentation patterns that are largely in agreement with the 70 eV patterns as available from the NIST database (see footnote 1). Relatively small inconsistencies in these patterns are likely due to the thermal codesorption of trapped H$_2$CO or CH$_3$OH with the aforementioned COMs, which have the same $m/z$ values for some of the dissociative ionization fragments.

As mentioned before, these COMs could not be detected unambiguously by means of RAIRS, because there exists considerable overlap between the strongest IR absorption features of methyl formate, glycolaldehyde, and ethylene glycol with those of H$_2$CO and CH$_3$OH that have higher abundances in the ice. Furthermore, an attempt to detect these COMs through the molecule specific but weaker C–C stretching mode of glycolaldehyde and ethylene glycol or the O–CH$_3$ stretching mode of methyl formate were unsuccessful due to the low final yield of these species. Nevertheless, important information can be derived from spectroscopic data. Fig. 4 presents RAIR spectra obtained at 15 K after 850 min of H$_2$CO hydrogenation with excess of atomic hydrogen in a H$_2$CO:H = 1:30 ratio and compared with the results from an experiment in which H$_2$CO and CH$_3$OH are deposited in a 1:1 ratio (exps 1.11 and 2.6). The two left-hand panels show zoom-ins of wavelength do- mains that cover the strongest absorption features of HCO and $CH_2OH$. The two right-hand panels coincide with the strongest ab- sorption features of glycolaldehyde. Moreover, a RAIR spectrum obtained after 1ML deposition of pure glycolaldehyde, as reported by Öberg et al. (2009), is included for a direct comparison. From this, it becomes clear that while the strongest absorption features of HCO and $CH_2OH$ cannot be seen (left-hand panel), those of $HC(O)CH_2OH$ can be observed in the ice, despite substantial overlap with $H_2CO$ and $CH_3OH$ features. For example, the $\sim 1748\,cm^{-1}$ peak of glycolaldehyde is positively identified in the left shoulder of the $H_2CO(\nu_2)$ band obtained after codeposition of $H_2CO$ with H-atoms, since this absorption band is not present in the spectra of pure $H_2CO:CH_3OH$ ice mixtures. Similarly, no contradiction is found with the possible presence of the $\sim 1112\,cm^{-1}$ absorbance feature of $HC(O)CH_2OH$ on the spectrum obtained in the $H_2CO+H$ experiment. This provides a strong argument that glycolaldehyde is formed already at 15 K among other products in a $H_2CO+H$ code- position experiment, and that detection of COMs during the TPD by means of the QMS is not the result of recombination of radicals trapped in the lattice of the ice at higher temperatures.

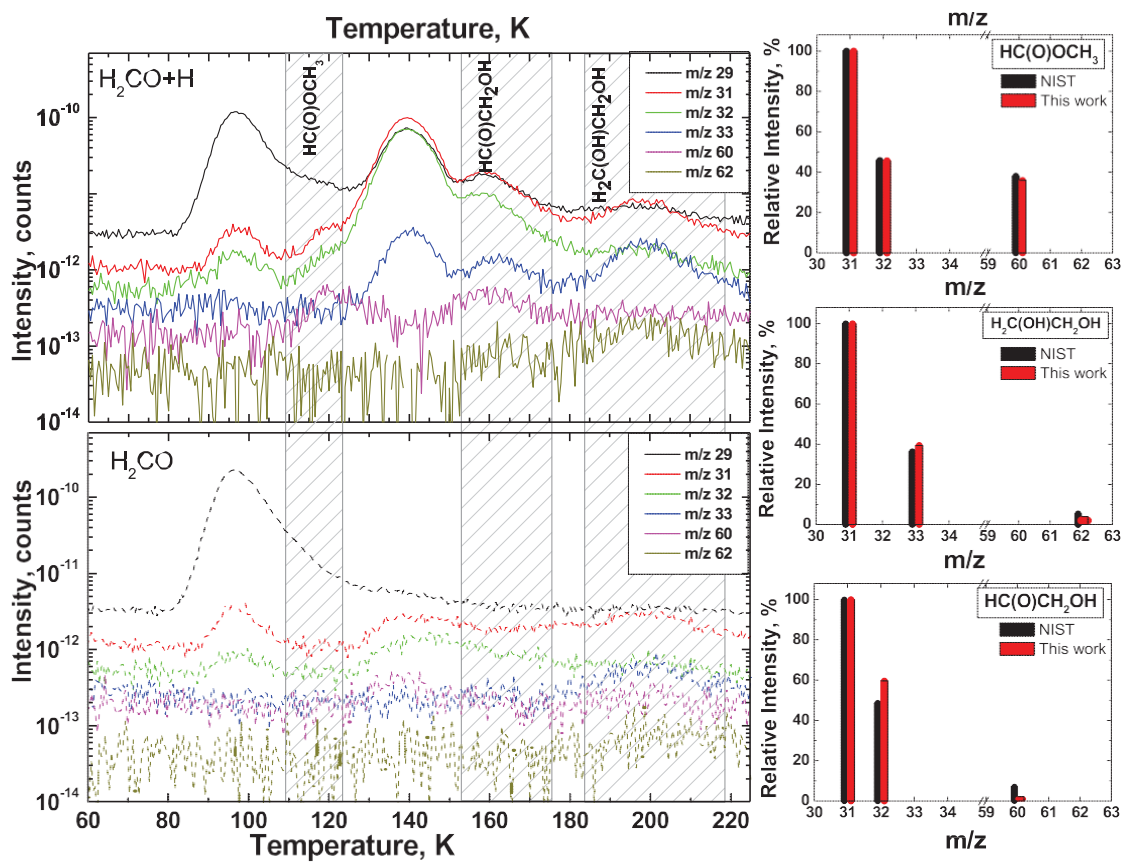

**Figure 3.** Left: the TPD mass spectra obtained after deposition of $H_2CO$ (lower panel) and equal amounts of $H_2CO$ and H-atoms (upper panel). The codepositions are performed at 15 K for 360 min, using a mixture of $H_2CO:H = 1:30$ and an H-atom flux equal to $6 \times 10^{14}$ atoms $min^{-1}\,cm^{-2}$. Only relevant $m/z$ numbers are shown. Right: comparison of the fragmentation patterns of the detected desorbing COMs with available literature values.

Furthermore, repeating the H$_2$CO codeposition with H-atoms at higher temperatures, i.e. at 25 and 50 K, does not result in COM detections. This is not surprising since the H-atom lifetime on the surface drops significantly at higher temperatures, which results in a drastic decrease of reactivity of H$_2$CO with H-atoms at 25 K and a complete inhibition at 50 K. A similar decrease of reaction rates with temperature was observed by Fuchs et al. (2009) and Hidaka, Kouchi & Watanabe (2007) upon CO hydrogenation.

### 3.4 Establishing the types of involved intermediate radicals

In an attempt to reveal in more detail the reaction mechanisms responsible for the formation of methyl formate, glycolaldehyde and ethylene glycol, several additional experiments are performed. Since different kinds of intermediate radicals are formed upon H- atom exposure of CO and CH$_3$OH, i.e. HCO and possibly CH$_2$OH (Nagaoka, Watanabe & Kouchi 2007), two-component binary mixtures of CO with H$_2$CO, CO with CH$_3$OH, and H$_2$CO with CH$_3$OH are codeposited with H-atoms at 15 K (exps 2.1, 2.3 and 2.5). Subsequently, relative COM abundances are examined in each of the experiments (see Table 1). To facilitate direct comparisons among these experiments, a 1:1 ratio is used for the molecular constituents, while the total applied ratio, including H-atoms, amounts to 1:1:30 to guarantee that (de)hydrogenation effects will be clearly visible.

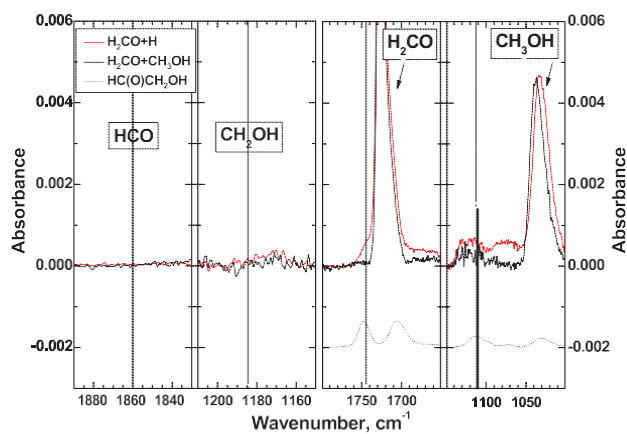

**Figure 4.** RAIR spectra obtained after codeposition of H$_2$CO and H-atoms for 850 min at 15 K using a ratio H$_2$CO:H = 1:30 (exp. 1.11) and codeposition of H$_2$CO with CH$_3$OH using a 1:1 ratio (exp. 2.6). The two left-hand panels show the regions of the spectra where the most intense absorption features of HCO and CH$_2$OH are expected. The two right-hand panels show the spectral regions where the strongest absorption features from HC(O)CH$_2$OH are present, scaled for the amount of H$_2$CO and CH$_3$OH, respectively. A spectrum of pure HC(O)CH$_2$OH (blue) is presented for comparison.

The formation of all three COMs is observed in each of the experiments with the exception of CO+CH$_3$OH+H, where methyl formate cannot be detected. Fig. 5 shows the comparison between the H$_2$CO+H and CO+CH$_3$OH+H experiments for selected TPD QMS spectra and the relevant $m/z$ values. In the H$_2$CO+H code- position (left-hand panel), a desorption peak centred around 120 K is clearly seen and assigned to HC(O)OCH$_3$ according to the QMS fragmentation shown in Fig. 3. However, in the CO+CH$_3$OH+H ice mixture experiment (right-hand panel), there are no mass signals that can be assigned to methyl formate. This is an important finding which shows that only the abundant presence of H$_2$CO in the sample produces HC(O)OCH$_3$ molecules.

Although some amount of H$_2$CO can be formed in the CO+CH$_3$OH+H experiment by hydrogenation of CO molecules and by dehydrogenation of CH$_3$OH (see signal 29 $m/z$ at $\sim$96 K in Fig. 5), this is not effective enough for our experimental settings to be transformed to a detectable amount of HC(O)OCH$_3$.

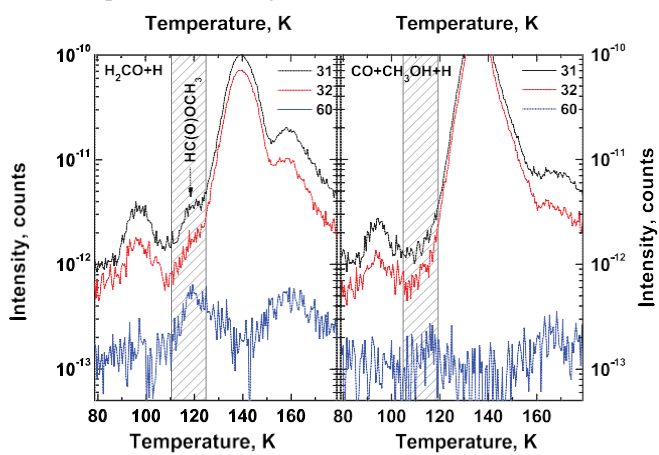

**Figure 5.** Left: the TPD QMS spectrum obtained after codeposition of H$_2$CO with H-atoms at 15 K for 360 min with H$_2$CO:H = 1:30. Right: TPD QMS spectrum obtained after codeposition of CO, CH$_3$OH and H- atoms with CO:CH$_3$OH:H = 1:1:30 for the same experimental conditions. The H-atom flux in both experiments is equal to $6 \times 10^{14}$ atoms min$^{-1}$ cm$^{-2}$. Only selected $m/z$ values from the desorbing species are indicated. The desorption peak centred around 120 K and assigned to HC(O)OCH$_3$ is visible in the left-hand panel, while it is absent among the products of the experiment shown in the right-hand panel. The 140 and 160 K peaks are due to methanol and glycolaldehyde, respectively.

### 3.5 Control experiments

H$_2$CO is sensitive to its ice surrounding, and is chemically active upon thermal processing, when embedded in an H$_2$O, CH$_3$OH, and/or NH$_3$ environment (Schutte, Allamandola & Sandford 1993; Duvernay et al. 2014). Also, despite cleaning procedures, low- level pollutions may be involved that influence final outcomes. Therefore, in order to exclude any artefacts, three control experiments (exps 3.1–3.3) involving the deposition of different H$_2$CO:CH$_3$OH ice mixtures (CH$_3$OH/H$_2$CO ratio = 0.1, 0.3 and 3.5) are performed and TPD QMS spectra are acquired. In Fig. 6, the normalized integrated QMS signals are shown for each of the three COMs found here and for each of the used mixing rates. In the same figure, also the COM intensities are plotted as obtained in our hydrogenation experiments versus the final H$_2$CO and CH$_3$OH abundance ratio obtained after 6 h of codeposition. It should be noted that the ethylene glycol, glycolaldehyde and methyl formate abundances presented in Fig. 6 are normalized with respect to the total amount of H$_2$CO and CH$_3$OH observed by RAIRS at the end of a codeposition experiment but before starting a TPD QMS experiment. The error bars represent instrumental errors and do not account for uncertainties resulting from the baseline subtraction procedure. The CO abundance is not taken into account here due to the expected chemical inertness of this species during thermal processing.

The comparison between hydrogenation and control experiments indicates that COMs are formed in the ice and cold surface H-atom addition and abstraction reactions are required to explain the observed COM abundances because the relative intensity of the newly formed COMs generally exceeds the one from the control experiments. However, there are some observations that need to be pointed out. For instance, in the case of H$_2$C(OH)CH$_2$OH (top panel of Fig. 6), some H$_2$C(OH)CH$_2$OH is present in the control experiments and a clear growing trend with increasing CH$_3$OH is observed and fitted with a polynomial function. This can be explained by either the presence of ethylene glycol as a contamination in the CH$_3$OH sample or by thermally induced chemistry involving CH$_3$OH molecules. However, the intensity of the observed H$_2$C(OH)CH$_2$OH produced by hydrogenation is significantly higher than that found in the control experiments for all hydrogenated ice mixtures except that with the highest CH$_3$OH abundance (exp. 2.3). For both HC(O)CH$_2$OH and HC(O)OCH$_3$ molecules (middle and bottom panels of Fig. 6), the observed intensities in the hydrogenation experiments cannot be reproduced in the control experiments, with the exception of the CO+CH$_3$OH+H and CO+H co-deposition experiments (exps 2.3 and 2.8), where the abundance of HC(O)OCH$_3$ is lower than the value obtained in the control experiments at the corresponding CH$_3$OH/H$_2$CO ratio.

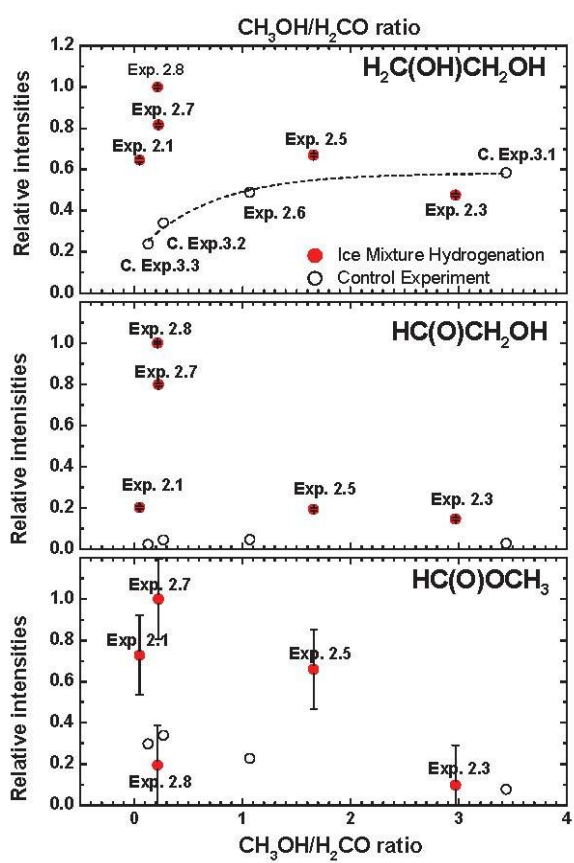

**Figure 6.** Normalized integrated TPD QMS intensities for each of the three COMs discussed here (upper panel = ethylene glycol, middle panel = glycolaldehyde and lower panel = methyl formate) as a function of the CH$_3$OH/H$_2$CO mixing ratio. The empty circles indicate the results of the control experiments, i.e. without hydrogenation. Solid circles show the result from the ice mixture hydrogenation experiments. The mass peaks used for the integration are $m/z$ = 31, 31 and 60 for upper, middle and lower panel, respectively. The numbers are normalized with respect to the total amount of H$_2$CO and CH$_3$OH observed before the TPD experiment.

To further understand the relative abundance of the produced COMs in all the performed experiments, the final COM abundances obtained after each experiment is normalized to the total amount of deposited carbon-bearing species, i.e., CO, H$_2$CO and CH$_3$OH. Results are presented in Fig. 7 (black columns). Since a non-negligible amount of H$_2$C(OH)CH$_2$OH is observed in all the control experiments (top panel of Fig. 6), this has been taken into account and subtracted from the final H$_2$C(OH)CH$_2$OH abundances observed in the hydrogenation experiments using a procedure described below. The control experiment data points (upper panel of Fig. 6) are fitted with the polynomial function. Subsequently, the obtained coefficients – 'H$_2$C(OH)CH$_2$OH$_{(control\ exp.)}$'/'H$_2$C(OH)CH$_2$OH$_{(hydrogenation\ exp.)}$' – are derived for all of the hydrogenation experiment data points. Consequently, the H$_2$C(OH)CH$_2$OH abundances are reduced by the obtained coefficients and presented in Fig. 7 (red columns in top panel of Fig. 7). The latter should be treated as a relative comparison of the lower formation limits of the observed COMs.

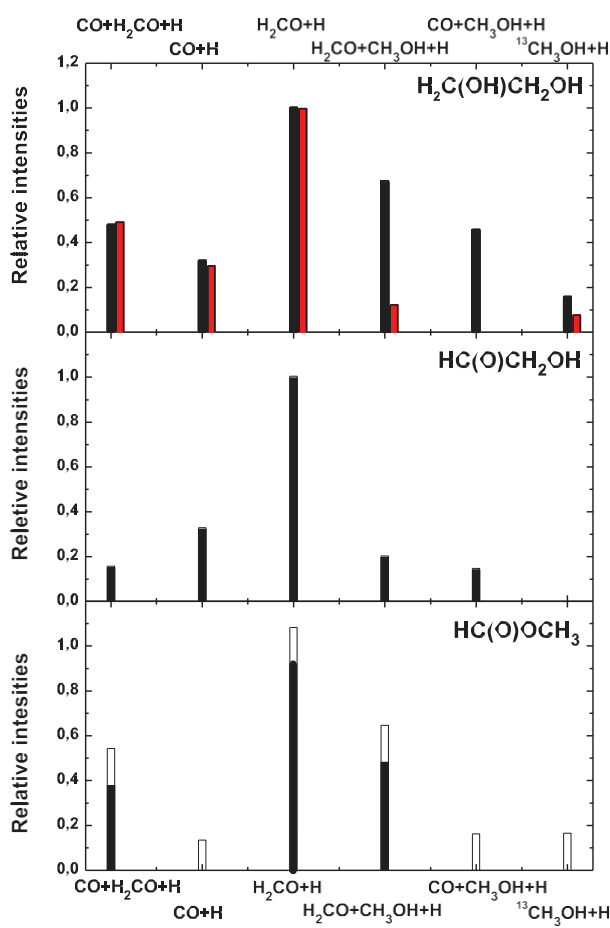

**Figure 7.** Relative comparison of the integrated intensities of ethylene glycol (upper panel), glycolaldehyde (middle panel), and methyl formate (lower panel) observed by means of TPD QMS and normalized for the total amount of carbon-bearing molecules observed before performing the TPD experiment. Black columns represent raw data. In the upper panel, red columns represent the data obtained after reduction of the control experiments shown in Fig. 6 and should be treated as relative comparison of the lower formation limits observed in this study. White bars (mainly visible in the lower panel, but present in all the panels) represent instrumental errors.

# 4 DISCUSSION

## 4.1 Chemical network

As described by Fedoseev et al. (2015), the formation COMs observed here can be explained by interaction of intermediate radicals that are formed upon H-atom addition and abstraction reactions with $H_2CO$. These are $CH_3O$ and possibly $CH_2OH$ radicals formed in the reaction:

$$H_2CO + H \rightarrow CH_3O \qquad (1a)$$

$$H_2CO + H \rightarrow CH_2OH, \qquad (1b)$$

which then further yield methanol through the reaction:

$$CH_3O + H \rightarrow CH_3OH, \qquad (2a)$$

$$CH_2OH + H \rightarrow CH_3OH, \qquad (2b)$$

and HCO radicals formed through the reaction:

$$H_2CO + H \rightarrow HCO + H_2, \qquad (3)$$

which is a necessary step to form CO through the reaction:

$$HCO + H \rightarrow CO + H_2. \qquad (4)$$

Various reactions involving these intermediate radicals can directly yield glycolaldehyde and ethylene glycol following reaction (1b):

$$HCO + CH_2OH \rightarrow HC(O)CH_2OH, \qquad (5)$$

$$CH_2OH + CH_2OH \rightarrow H_2C(OH)CH_2OH, \qquad (6)$$

or methyl formate in case of reaction (1a):

$$HCO + CH_3O \rightarrow HC(O)OCH_3. \qquad (7)$$

This scheme is also fully consistent with the formation route of ethylene glycol and glycolaldehyde through the sequence of reactions involving glyoxal proposed by Fedoseev et al. (2015):

$$HCO + HCO \rightarrow HC(O)CHO, \qquad (8)$$

$$HC(O)CHO + 2H \rightarrow HC(O)CH_2OH, \qquad (9)$$

$$HC(O)CH_2OH + 2H \rightarrow H_2C(OH)CH_2OH. \qquad (10)$$

The presence of methyl formate among the products observed in this study indicates that reaction (1a,b) should at least partially result in the formation of $CH_3O$ (instead of $CH_2OH$) to yield $HC(O)OCH_3$ through reaction (7). Methyl formate was not observed by Fedoseev et al. (2015); the use of CO as a starting point in that study resulted in considerably lower final yields of $H_2CO$ and $CH_3OH$ than in this work and, therefore, significantly lower amounts of formed $CH_3O$ than required to yield methyl formate.

Interaction of $H_2CO$ molecules with H-atoms results in both the formation of HCO and $CH_3O$ radicals via H-atom abstraction and addition, respectively. The latter process can also lead to the formation of $CH_2OH$ radicals in the case that the H-atom addition takes place on the oxygen side of the $H_2CO$ molecule. The non-detection of $HC(O)OCH_3$ among the detected COMs in $CO+CH_3OH+H$ implies that one of these radicals must play a crucial role in the formation of methyl formate. This cannot be HCO, as it can be produced by H-atom additions to CO molecules present in the ice mixture, thus it must be a reaction product of H-atom additions to $H_2CO$. Taking into account reaction (7), which describes $HC(O)OCH_3$ formation, this radical should be $CH_3O$. This, in turn, allows drawing an important conclusion: H-atom abstraction reactions from $CH_3OH$ do not result in efficient formation of $CH_3O$ radicals but yield primarily $CH_2OH$:

$$CH_3OH + H \rightarrow CH_2OH + H_2, \qquad (11)$$

as otherwise the presence of $H_2CO$ would not be a pre-requisite. This conclusion is consistent with the result of Nagaoka et al. (2007) and Hidaka et al. (2009) obtained by studying deuterium substitution in methanol. Furthermore, H-atom addition reactions to $H_2CO$ should result in considerable amounts of formed $CH_3O$ through reaction (1), while the formation of $CH_2OH$ in this reaction is expected to be a less efficient process.

The non-production of glycolaldehyde in the $CO+CH_3OH+H$ codeposition experiment (exp. 2.3) obtained after subtraction of the control experiment data should be stressed here. This suggests that the $H_2C(OH)CH_2OH$ formation mechanism through reactions (11) and (6) is overall less efficient than through reactions (5) and (10) or reactions (8)–(10). This may be an indication of a lower formation rate of $CH_2OH$ in reaction (11) compared to the HCO formation rates in $CO + H$ or in reaction (3).

Another reason why recombination of $CH_2OH$ radicals to yield $H_2C(OH)CH_2OH$ seems overall less efficient than recombination of HCO with HCO, $CH_3O$ or $CH_2OH$ radicals may be due to a geometrical properties of the species involved. The access to the unpaired electron of $CH_2OH$ radical is significantly blocked by H-atoms bonded to carbon and oxygen atoms, while in the case of HCO radical the access to the unpaired electron will be easier. Thus, one can expect that the rate of $CH_2OH$ radical recombination is less probable, or, alternatively, results in H-atom abstraction to form methanol and formaldehyde:

$$CH_2OH + CH_2OH \rightarrow CH_3OH + H_2CO, \qquad (12)$$

due to the easier access of H-atoms to C-H bonds.

The proposed COM formation network based on all investigated reaction routes is presented in Fig. 8. From top to bottom, a chain of H-atom addition and abstraction reactions leading to the formation of $CH_3OH$ from CO is shown. As confirmed in this study, $H_2CO$ can undergo an abstraction reaction induced by H-atoms to form HCO radicals, which successively can be dehydrogenated to form simple molecules, i.e. CO, thus increasing the total number of HCO formation events and its lifetime in the ice mantle. $H_2CO$ also participates in addition reactions with H-atoms. Formation of the $CH_3O$ radical is confirmed in this study by observing methyl formate; however, formation of $CH_2OH$ radicals cannot be excluded, since all experiments involving $H_2CO$ codeposition demonstrate relatively high yields of glycolaldehyde and ethylene glycol. In contrast, H-atom- induced abstraction reaction involving $CH_3OH$ likely yield $CH_2OH$ while no proof for $CH_3O$ formation is found.

The barrier-less recombination of HCO intermediates yielding glyoxal followed by consequent hydrogenation, i.e. the mechanism investigated in Fedoseev et al. 2015, yields $HC(O)CH_2OH$ and $H_2C(OH)CH_2OH$ and is presented in the right-hand panel of the diagram. Alternatively, the intermediate HCO radicals can directly recombine with $CH_3O$ or $CH_2OH$ to form $HC(O)OCH_3$ and $HC(O)CH_2OH$, respectively. $CH_2OH$ and $CH_2OH$ recombination (dash arrows) seems to contribute less to the formation of $H_2C(OH)CH_2OH$. This may be explained by geometrical constraints or overall low efficiencies of abstraction reactions involving methanol.

From our experimental results, we cannot confidently determine whether diffusion of the intermediate radicals is involved in the formation of the observed methyl formate, ethylene glycol and glycolaldehyde at 15 K. In the simulations reported by Fedoseev et al. (2015), high activation barriers are used for HCO and $CH_3O$ diffusion. This effectively immobilizes such species and only radicals formed next to each can recombine. Fedoseev et al. (2015) showed that the formation and reaction of adjacent radicals ex- plains the observed results, offering an efficient formation pathway even at the low temperatures typical for dense dark clouds (∼10 K). This is consistent with the laboratory detection of the three COMs discussed here. It should be noted that a similar reasoning is often used to explain results obtained in photoprocessing experiments, i.e. two consequent photodissociation events result in the formation of radicals that can recombine when located next to each other in the bulk of the ice (Öberg et al. 2009).

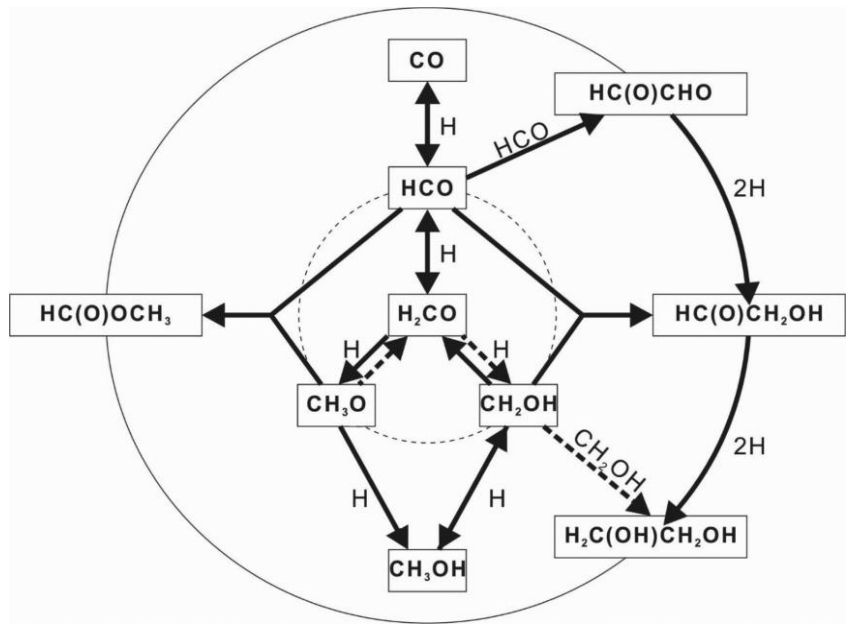

**Figure 8.** Extended COM formation network as obtained from the CO, $H_2CO$, and $CH_3OH$ hydrogenation experiments. Solid arrows indicate the reaction pathways confirmed or suggested in this study. Dashed lines indicate the overall less efficient pathways.

Another mechanism that may be important is the diffusion of newly formed intermediate radicals upon exothermic formation. The involved excess energy supports the diffusion of molecules and radicals even at low temperature. Radicals produced upon photodissociation in the ice initially have excess energy and can diffuse, as demonstrated by molecular dynamics simulations (*e.g.* Andersson et al. 2006). From our measurements, the impact of this process cannot be confirmed.

### 4.2 Astrochemical implications and conclusions

In dense molecular clouds, the reaction between CO molecules and H-atoms accreting on the grain surface does not only explain the abundance of interstellar methanol (Watanabe et al. 2006; Cuppen et al. 2009), but also its deuterium enrichment (Nagaoka et al. 2005; Hidaka et al. 2009). The latter can be realized through the series of H-atom abstraction reactions from the C-H ends of $H_2CO$ and $CH_3OH$ followed by D-atom additions, which take place faster than the corresponding D-atom abstraction reactions followed by substitution with H-atoms. HCO and $CH_2OH$ are possible intermediates in these H-atom abstraction reactions.

In this work, we confirm that such H-atom-induced abstraction reactions take place by experimental observations of $H_2CO$ formation upon H-atom exposure of $CH_3OH$ and CO formation upon H-atom exposure of $H_2CO$ for an astronomically relevant temperature of 15 K. This means that a CO hydrogenation mechanism leading to the formation of interstellar methanol in dark molecular clouds is not irreversible. $CH_3OH$ as well as its formation intermediates can participate in sequences of consecutive H-atom addition and abstraction reactions, thus, once formed, $H_2CO$ and $CH_3OH$ molecules can be potentially dehydrogenated and become once again available for hydrogenation.

Interactions between reactive radicals such as those produced here are actively discussed in the literature as a source for COM formation, both theoretically and experimentally (Hudson & Moore, 2000; Bennett et al. 2007; Garrod et al. 2008; Öberg et al. 2009; Woods et al. 2012; Garrod 2013). In these models, UV photon or cosmic-ray-induced dissociation of $CH_3OH$ is usually taken as the external trigger responsible for the formation of these various intermediates. However, Fedoseev et al. (2015) experimentally showed that glycolaldehyde ($HC(O)CH_2OH$) and ethylene glycol ($H_2C(OH)CH_2OH$) can be equally efficiently formed just by cold surface hydrogenation of CO molecules without involvement of UV- or cosmic-ray energetic processing of interstellar ices at 15 K. Here, we report that in addition to glycolaldehyde and ethylene glycol formation also methyl formate ($HC(O)OCH_3$) production is observed. The suggested mechanism of $HC(O)OCH_3$ formation is the radical–radical recombination of HCO, formed either by H-atom addition to a CO molecule or H-atom abstraction from $H_2CO$, and $CH_3O$ produced by H-atom addition to $H_2CO$. Furthermore, along with the formation of glycolaldehyde and ethylene glycol through the recombination of HCO radicals and the subsequent hydrogenation of glyoxal (Woods et al. 2013; Fedoseev et al. 2015), glycolaldehyde can be formed through the direct recombination of HCO and $CH_2OH$ radicals, while ethylene glycol can form through the recombination of two $CH_2OH$ radicals. However, the latter radical- radical reaction is found to be less efficient than the hydrogenation of glycolaldehyde under our experimental conditions.

Codeposition of $H_2CO$ with H-atoms at 25 and 50 K results in a decrease of the efficiency of both abstraction and addition reactions; consequently, no formation of COMs is observed at these temperatures. This can be explained by a substantial drop in the life-time of H-atoms on the ice surface with increase of the temperature.

The direct consequence of both H-atom addition and abstraction reactions is to increase the number of interaction events as well as the timespan over which radicals reside in the ice. This should increase the overall reactivity and likely more COMs are formed through recombination of reactive intermediates than assumed so far. Clearly, the seemingly opposite processes of H-atom addition and abstraction reactions will decrease the overall efficiency to form methanol directly from CO hydrogenation, essentially shifting the equilibrium point. The rates inferred by Fuchs et al. (2009) should therefore, be regarded as effective rates. In parallel, other processes become possible, increasing the overall efficiency with which COMs are formed. This process that is studied here only for a few temperature settings, is expected to be temperature dependent. The main take-home message from this work is that addition and abstraction reactions upon H-atom exposure of ice mantles can ex- plain the formation of COMs in dense molecular clouds even when energetic external UV radiation or cosmic rays are lacking. It also means that solid state COM formation can start already at the be- ginning of the CO freeze-out stage, well before $CH_3OH$ containing ices are thermally and energetically processed by the heating and radiation of the emerging protostar. It should be mentioned, that diffusion related processes will be different for the long astrochemical time-scales at play and can enhance the overall efficiency in comparison with the short laboratory time-scales. This provides further support of the COM formation mechanisms discussed in this work. This has important implications in astrobiology; glycolaldehyde is the simplest representative of the aldoses family to which sugars like glycose, ribose and erythrose belong, while ethylene glycol is the simplest polyol among which the triol glycerin is well known. As such, the non-energetic processes discussed here provide an important alternative to the formation of these prebiotically relevant species at an early stage in the chemical evolution of dark interstellar clouds.

The recent gas phase detection of COMs in pre-stellar cores, i.e., environments where temperatures are too low to initiate thermal desorption, raises questions concerning the efficiency of solid state formation of complex molecules and the process(es) responsible for their desorption. Typically, non-thermal desorption mechanisms, that is, upon impacting cosmic rays or irradiation by secondary UV photons, are expected to explain the effective transfer from solid state to gas phase (Bacmann et al. 2012; Cernicharo et al. 2012). In the case of CO, non-dissociative photodesorption explains the observed gas phase abundances (Fayolle et al. 2011), but for other species, like methanol, photodissociation seems to offer a competing scenario, (Öberg et al. 2009). However, larger species may be able to dissipate excess energy more effectively, due to the larger amount of vibrational modes that will help decreasing the dissociation efficiency. Another desorption mechanism that may be relevant is through cosmic ray induced impulsive spot heating. This process has been descripted in detail by Ivlev et al. (2015). The model presented by Garrod, Wakelam & Herbst (2007) shows that chemisorption offers another alternative mechanism; excess energy due to exothermicity of surface reactions offers a low temperature non-thermal desorption pathway. It is possible, given the nature of the reactions discussed in this work that COMs formed in CO- rich ices experience chemisorption. Moreover, intermolecular (van der Waals like) interactions with CO will be weak, compared to hydrogen bonds in water rich ices. At the moment, the nature of the process bridging the grain–gas gap is still unclear, and the work presented here offers good arguments for a further focus in future studies.

**ACKNOWLEDGEMENTS**

This research was funded through a VICI grant of NWO, the Netherlands Organization for Scientific Research, NOVA (the Netherlands Research School for Astronomy), A-ERC grant 291141 CHEMPLAN and the FP7 ITN LASSIE (GA 238258). Furthermore, SI acknowledges the Royal Society for financial support.